\documentstyle[12pt]{article}

\newcommand{\be}{\begin{equation}}
\newcommand{\ee}{\end{equation}}
\newcommand{\ba}{\begin{eqnarray}}
\newcommand{\ea}{\end{eqnarray}}
\newcommand{\bea}{\begin{eqnarray}}
\newcommand{\eea}{\end{eqnarray}}

\newcommand{\cV}{\langle {\cal V}|}

\newcommand{\non}{\nonumber\\}
\newcommand{\eq}[1]{(\ref{#1})}

\newcommand{\cH}{{\cal H}}
\newcommand{\hcH}{{\widehat{\cal H}}}
\newcommand{\Nup}{{N^{(\uparrow)}}}
\newcommand{\Ndown}{{N^{(\downarrow)}}}
\newcommand{\Mup}{{M^{(\uparrow)}}}
\newcommand{\Mdown}{{M^{(\downarrow)}}}

\newcommand{\bZ}{{\bf Z}}
\newcommand\bC{\bf C}

\newcommand{\isum}[3]{\sum_{#1=#2}^{#3}}

\newcommand{\calH}{{\cal H}}

\newcommand{\calHa}{\widehat{{\cal H}}}

\newcommand{\phii}[1]{\phi^{(#1)}}

\newcommand{\boson}[2]{a_{#1}^{(#2)}}

\newcommand{\si}{\sigma}
\newcommand{\sj}{\sigma'}

\newcommand{\jack}[2]{J_{{#1}}^{({#2})}}
\newcommand{\del}[1]{D_{#1}}

\newcommand{\vertex}[4]{%
\Gamma(#1,#2,\cdots,#3;#4)
}

\renewcommand{\a}{\alpha}
\newcommand{\Pu}[1]{p^{(1)}_#1} 
\newcommand{\Pd}[1]{p^{(2)}_#1} 
\newcommand{\Mat}[1]{\left[\matrix{#1}\right]}

\begin{document}
\begin{titlepage}
\nopagebreak
%
%

\begin{flushright}
December 1995\hfill YITP/95-20\\
hep-th/9512065
\end{flushright}

\renewcommand{\thefootnote}{\fnsymbol{footnote}}
\vfill
\begin{center}
{\Large Collective Field Description of}
\vskip 2mm
{\Large Spin Calogero-Sutherland Models}

\vskip 20mm

{\large H.~Awata\footnote{JSPS fellow}\footnote{
E--mail: awata@yukawa.kyoto-u.ac.jp},
Y.~Matsuo\footnote{E--mail: yutaka@yukawa.kyoto-u.ac.jp},
T.~Yamamoto\footnote{E--mail: yam@yukawa.kyoto-u.ac.jp}
}
\renewcommand{\thefootnote}{\arabic{footnote}}
\vskip 10mm
{\em Yukawa Institute for Theoretical Physics}\\
{\em Kyoto University}\\
{\em Kitashirakawa, Kyoto 606-01, Japan}
\end{center}
\vfill

\begin{abstract}
Using the collective field technique,
we give the description of the spin
Calogero-Sutherland Model (CSM) in terms of free bosons.
This approach can be applicable
for arbitrary coupling constant
and provides the bosonized Hamiltonian of the spin CSM.
The boson Fock space can be identified with
the Hilbert space of the spin CSM in the large $N$ limit.
We show that the eigenstates corresponding
to the Young diagram with a single row or column
are represented by the vertex operators.
We also derive a dual description
of the Hamiltonian and comment on the construction
of the general eigenstates.
\end{abstract}

\vfill
\end{titlepage}

%
%
\section{Introduction}
The Calogero-Sutherland model (CSM) \cite{C,S}
has been an interesting laboratory to study the
fractional statistics in $(1+1)$-dimension \cite{EX,BW,Ha}.
Its paradigmatic r\^ole as the anyonic analog of the free
boson or fermion gas has been established.
Also, the CSM is related to various branches of physics
and contains many interesting aspects
in mathematical physics \cite{revCS}.
Especially, it is known that this model is
the universal Hamiltonian for the disordered systems \cite{MIT}.

Many variants of the CSM now exist,
for example, its lattice cousin,
the so-called Haldane-Shastry model \cite{HS},
and multicomponent version,
the {\it spin} (or dynamical) CSM \cite{HaHa,KaKu,MP,HW}.
A lot of intriguing results have been obtained in connection with
these models where
the Yangian symmetry \cite{HHTBP,BGHP,Hikamia,BHW}
plays essential r\^ole to explain the degeneracy of the spectrum.
For particular couplings, $\alpha=2,1/2$,
this nonlinear symmetry is known to be realized
through the spinon basis
(or the vertex operators of the free boson) \cite{BPS,BLS,AN}.
This is the point where the symmetry of the system is enhanced
to the level one $su(2)$ Kac-Moody algebra.

In our previous studies \cite{AMOS},
the bosonization for the CSM has been given
(see for the related works \cite{MY,Iso,MPb,Khv}).
One of the essential observations in those works
was that the collective coordinate description of the system
is equivalent to
the Coulomb gas description of the minimal model of
conformal field theory. In particular, two screening currents
of the minimal model are naturally identified with the generating
functionals of one particle and one hole states.
Similarly, any eigenstate (which is known as the Jack polynomial)
can be identified with the singular vector of the appropriate $W$
algebra.

In this letter, we show that some part of the above scenario can be
generalized to the spin CSM
without any restriction on the coupling constant.
We describe the Hamiltonian in terms of
multicomponent free bosons.
In our method, the correspondence between the
spin CSM Hilbert space and the free boson Fock space is one to one.
We explicitly obtain the generating functional
of one particle (hole) excited states as vertex operator.
General eigenstates would be written as the product of the
vertex operators.
We also derive the ``dual'' Hamiltonian
defined by the action of the original Hamiltonian on such states.
The integrability of this dual Hamiltonian
directly follows from its construction.

There are, however, some differences from the spinless CSM.
For example,
the duality
(or the charge conjugation) symmetry of the system disappears.
Therefore, it becomes rather difficult to relate
the Hamiltonian with the loop algebra such as the Virasoro algebra
or the Kac-Moody algebra.
In particular, it is still hard to see the connection with
the Yangian or the Kac-Moody symmetry
even if we pick $\alpha=2$ or $1/2$.

In the conclusion, we comment how one can
construct the general eigenstates of the spin CSM
by using the dual Hamiltonian.

\section{Collective Field Description
of Spin CSM}

Let us write down the reduced form of
the Hamiltonian for the spin CSM (see \cite{KaKu,BGHP}).
Performing a ``gauge'' transformation,
the Hamiltonian is given by
\be
\cH = \alpha \sum_{i=1}^N D_{x_i}^2
+\sum_{i<j} \frac{x_i+x_j}{x_i-x_j}(D_{x_i}
-D_{x_j})-2
\sum_{i<j} \frac{x_i x_j}{(x_i-x_j)^2} (1-K_{ij}),
\label{Hamil}
\ee
where
$\alpha\in \bC$ is the coupling constant,
$D_{x}\equiv x\frac{\partial}{\partial x}$
and $K_{ij}$ is the (coordinate) exchange operator,
namely, for a function $f$ in $x_i$'s,
$$K_{ij} f(\cdots,x_i,\cdots,x_j,\cdots) \equiv
f(\cdots, x_j, \cdots, x_i, \cdots).$$
The wave function of the Hamiltonian is described by
the coordinates $x_i$ and the spin variables $\sigma_i$ attached to
them. Each spin variable takes values in the set $\{ 1,2,\cdots,s\}$
({\it i.e.}, we consider the $s$-component system)
and the wave function should be invariant under
the simultaneous exchange of both variables,
\be
\psi(\cdots,x_i\sigma_i,\cdots,x_j\sigma_j,\cdots)
=\psi(\cdots,x_j\sigma_j\,\cdots,x_i\sigma_i,\cdots).
\ee
In other words, the exchange of the coordinate and that of the
spin variable have same effect when they are acted on the
wave function.


One of the nontrivial properties of the spin CSM Hamiltonian is that,
when we try to diagonalize the Hamiltonian,
we are able to restrict the Hilbert space
such that the spin variable for each particle is fixed.
More precisely, let us denote
$ x^{(\si)}_i$'s as those coordinates whose
spin variables take value $\si\in \{1,\cdots,s\}$.
Then the restricted Hilbert space is defined by
the set of functions which are symmetric under
the exchange $x^{(\si)}_i$ and $x^{(\si)}_j$ for each spin $\si$.
At first glance such restriction may not be
compatible with the action of the Hamiltonian
because of the terms which include the exchange operator $K_{ij}$.
These terms exchange the $x$ coordinates alone
and leave the spin variables untouched.
However, we can prove by an explicit computation
that the undesirable terms vanish.

In the restricted Hilbert space,
we can apply the standard bosonization
(or collective coordinates) technique.
Let us define the power sum for each spin,
\be
p^{(\si)}_n \equiv \sum_{i=1}^{N^{(\si)}} (x^{(\si)}_i)^n,
\ee
where $N^{(\si)}$ is the number of particles with spin $\si$.
We also introduce free bosons $a^{(\si)}_n$, $n\in\bZ$,
and boson fields $\phii{\si}_{\pm}(\xi),\ (\si=1,\cdots,s)$,
such that,
\be
\label{boson-fields}
[ \boson{n}{\si},\boson{m}{\sj} ]=n\delta^{\si,\sj}\delta_{n+m,0},
\qquad
\phi_\pm^{(\si)}(\xi)
=\mp\sum_{n>0}\frac{1}{n}\boson{\pm n}{\si}\xi^{\mp n}.
\ee
The bosonization method is to replace $p^{(\si)}_n$
by the free boson creation operator $a^{(\si)}_{-n}$.
The collective coordinate description becomes exact in
the limit that the number of the particles,
{\it i.e.}, $N^{(\si)}$'s become all infinite.
The replacement $p^{(\si)}_n\leftrightarrow a^{(\si)}_{-n}$
can be systematically carried out by
introducing the operator $\cV$,
\be
\cV
\equiv
\langle N| \,
\exp\left\{ \sum_{\si}\sum_{n>0}{\frac 1 n}
p^{(\si)}_n a^{(\si)}_n \right\},
\ee
with the lowest weight state $\langle N|$ such that
$\langle N| a^{(\si)}_{-n} = 0$, $n>0$ and
$\langle N| a^{(\si)}_0    = N^{(\si)} \langle N|$.
Taking the inner product with this bra state,
we can translate the Fock space of free bosons into
the restricted Hilbert space of the spin CSM.
Namely, $\cV$ translates
coordinates $x^{(\si)}$ to bosons $a^{(\si)}_n$
as follows,
\be
p^{(\si)}_n \cV = \cV a^{(\si)}_{-n},\qquad
n \frac{\partial}{\partial p^{(\si)}_{n}} \cV = \cV a^{(\si)}_n.
\ee
In the limit $N^{(\si)}\rightarrow \infty$,
this correspondence is one to one.  In other words,
any operator which acts on the restricted
Hilbert space can be rewritten by free boson oscillators.
In particular, the Hamiltonian is bosonized as follows.
Firstly, we shall decompose the Hamiltonian (\ref{Hamil})
into two parts,
\be
\cH(x) = \sum_{\si=1}^s \cH^{(\si)}(x^{(\si)}) +
\sum_{\si<\sj}\cH^{(\si\sj)}_{int}(x^{(\si)},x^{(\sj)}),
\ee
with
\ba
\cH^{(\si)}
& = & \alpha \sum_{i=1}^{N^{(\si)}}\left(D_{x^{(\si)}_i}\right)^2
+
\sum_{i<j}\frac{x^{(\si)}_i+x^{(\si)}_j}{x^{(\si)}_i-x^{(\si)}_j}
(D_{x^{(\si)}_i}-D_{x^{(\si)}_j}),
\\
\cH^{(\si\sj)}_{int} & = & \sum_{i,j}
\frac{x^{(\si)}_i+x^{(\sj)}_j}{x^{(\si)}_i-x^{(\sj)}_j}
(D_{x^{(\si)}_i}-D_{x^{(\sj)}_j})
-
2 \sum_{i,j}\frac{x^{(\si)}_i x^{(\sj)}_j}{(x^{(\si)}_i-x^{(\sj)}_j)^2}
(1-K_{x^{(\si)}_i,x^{(\sj)}_j}).
\label{decomp}
\ea
Then, the bosonized Hamiltonian
$\hat\cH
=
\sum_\si \hat\cH^{(\si)}+\sum_{\si<\sj}\hat\cH^{(\si\sj)}_{int}$,
where
$\cH \cV = \cV \hat\cH$,
is given by the formulae,
\ba
\label{bosonHamiltonian1}
\hat\cH^{(\si)} &=& \sum_{n,m>0}
\left( a^{(\si)}_{-n} a^{(\si)}_{-m} a^{(\si)}_{n+m} +
\alpha a^{(\si)}_{-n-m} a^{(\si)}_n a^{(\si)}_m \right)
+\sum_{n>0} \left(\alpha n - n + a^{(\si)}_0\right)
 a^{(\si)}_{-n} a^{(\si)}_n,
\\
\label{bosonHamiltonian2}
\hat\cH^{(\si\sj)}_{int} &=&
\sum_{n,m>0} a^{(\si)}_{-n} a^{(\sj)}_{-m}
\left(a^{(\si)}_{n+m} + a^{(\sj)}_{n+m}\right)
+\sum_{n>0}\left(a^{(\sj)}_0 a^{(\si)}_{-n} a^{(\sj)}_n
	       + a^{(\si)}_0 a^{(\sj)}_{-n} a^{(\si)}_n \right)
\non
&+&
\oint\frac{d\xi}\xi \frac{d\eta}\eta
\sum_{n,m\geq 0} \xi^{n} \eta^{m} a^{(\si)}_{-n} a^{(\sj)}_{-m}
\,e^{\sum_{n>0}{\frac 1n}(\xi^{-n}-\eta^{-n})(a^{(\sj)}_n - a^{(\si)}_n) }
\sum_{k>0}k \left( \frac{\xi^k}{\eta^k} +
 \frac{\eta^k}{\xi^k} \right).
\ea
Here $\oint \frac{dx}{x} f(x)$ stands for the constant term of $f(x)$.
The proof is similar to that in our previous papers \cite{AMOS}.
The essential point is that
$\cH^{(\si\sj)}_{int} \cV$ has no pole at $x^{(\si)}_i = x^{(\sj)}_j$
and is a power series in $x^{(\si)}_i$ and $x^{(\sj)}_j$'s.
To treat the parts which include the exchange operators, we used
\ba
K_{x^{(\si)}_i x^{(\sj)}_j} \cV
&=&
\cV
\,e^{\sum_{n>0}{\frac 1n}
\left((x^{(\si)}_i)^n - (x^{(\sj)}_j)^n\right)(a^{(\sj)}_n - a^{(\si)}_n) }
\non
&=&
\cV
\oint\frac{d\xi}\xi \frac{d\eta}\eta
\sum_{n,m\geq 0} \xi^{n} \eta^{m} (x^{(\si)}_i)^n (x^{(\sj)}_j)^m
\,e^{\sum_{n>0}{\frac 1n}(\xi^{-n}-\eta^{-n})(a^{(\sj)}_n - a^{(\si)}_n) }.
\ea

In the Appendix, we will give examples of
the eigenstates of this bosonized Hamiltonian for the low degree cases.

Remark that the third term of $\hat\cH^{(\si\sj)}_{int}$
is rewritten by using boson fields $\phi^{(\si)}_+(\xi)$ and
$D_\xi\phi^{(\si)}_{\leq0}(\xi) \equiv
\sum_{n>0}\boson{-n}{\si}\xi^{n} + a^{(\si)}_0$
as follows,
\be
\oint _{|\eta|>|\xi|}
{d\eta d\xi \over (\eta - \xi)^2}
:
D_\xi \phi^{(\si)}_{\leq0}(\xi) e^{\phi^{(\si)}_+(\xi)-\phi^{(\sj)}_+(\xi)}
D_\eta\phi^{(\sj)}_{\leq0}(\eta)e^{-\phi^{(\si)}_+(\eta)+\phi^{(\sj)}_+(\eta)}
:.
\ee
Here $:*:$ is the usual normal ordering.

%
%
\section{One Particle (Hole) States and Vertex Operators}
In this section, we will show that the wave function
of the one particle
(hole) excited states can be expressed as the vertex operator
of the free bosons.
Before proceeding to the explanation, it may be better to illustrate
the characterization of each eigenstate.
As it is well-known, the eigenstates of the CSM
without spin degrees of freedom can be
indexed by the Young diagrams.
Each row (column) in the diagram corresponds to the
particle (hole) excitations of the CSM (see, for example, \cite{Ha}).
For each diagram,
there is only one eigenstate, and the eigenvalue is
determined from the diagram.

Even if we introduce the
spin degrees of freedom, most of the structure remains
the same.  The eigenstates are again indexed by the Young diagrams.
The eigenvalue is also determined by the diagram and it
is actually the same as  the spinless case.
The difference, however,
is that the eigenstate is not unique for each diagram,
{\it i.e.}, the spectrum is degenerate.
This is caused by the existence of the
Yangian symmetry \cite{HHTBP,BGHP}.

There is a simple method to count the degeneracy of states for
each Young diagram.  With $s$ colors that we have, we paint each
box of the diagram according to the rule:
the boxes in the same row have the same color and
there is no constraint for the colors in the each column.
The colored Young diagram after this prescription
is indexed as $(\lambda_1\sigma_1,\lambda_2\sigma_2,
\cdots,\lambda_N\sigma_N)$
where $\lambda_i\in \bZ$ with $\lambda_1\geq\lambda_2\geq
\cdots\geq\lambda_N\geq 0$
and $\si_i\in\{1,\cdots,s\}$.
We identify the
diagrams which can be obtained from one another by
permuting the colors for each row with the same length.

This prescription is an obvious consequence of the fact
that the number of the boxes for each row can be identified with
the momentum of a quasi-particle. Since it has a color,
we need to paint each row by the same color. On the other hand,
the number of boxes
of each column is identified with the momentum of a quasi-hole.
The colors which appear on each column can be identified with
the colors of a quasi-particle which occupies the upper levels.
Let us illustrate it in Fig.1.
For simplicity we pick $\alpha=1$ and consider
the state depicted in Fig.1A.
There are three particles written as $a$, $ b$ and $ c$ with spin
1, 2 and 1, respectively, and four holes $x$, $y$, $z$ and $w$.
This state can be rewritten as the Young diagram in
Fig.1B. We see that particles are mapped to the rows
and holes to the columns, respectively.

\vskip 4mm
\input epsf.tex
\centerline{\epsfbox{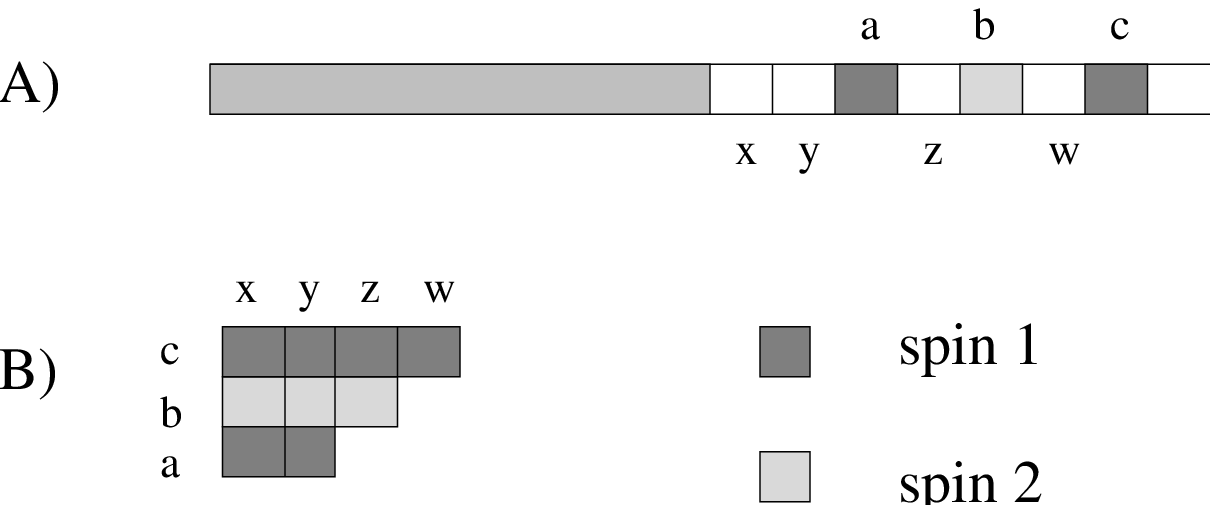}}
\centerline{{\bf Figure 1:}
A example of the colored Young diagram.}
{~}\vspace{1mm}

We are now in position to describe the vertex operator
construction of the eigenstates.
In the spinless situation \cite{AMOS},
we observed that only two types of the vertex operators,
$
\exp(\gamma \phi_-(\xi))
$
with $\gamma=1/\alpha,-1$,
have  ``simple'' forms after they are operated by
the Hamiltonian.  If we expand $e^{\phi_-(\xi)/\alpha}$
(resp. $e^{-\phi_-(\xi)}$) with respect to $\xi$, the
coefficient of $\xi^n$ is identified with the eigenstate
for the Young diagram $(n)$ (resp. $(1^n)$).
Even for the system with spin degrees of freedom,
we expect similar vertex operators
give the eigenstates indexed by diagrams with
a single row or a single column.

Let us introduce basic vertex operators,
\be
\label{vertex}
\Gamma(\xi;\gamma)
=
\cV
e^{\gamma \phi_-(\xi^{(1)})}\cdots e^{\gamma \phi_-(\xi^{(s)})}
|
N
\rangle
=
\exp
\left[
\gamma\sum_{\si=1}^s\sum_{n=1}^{\infty}\frac{1}{n}p_n^{(\si)}(\xi^{(\si)})^n
\right].
\ee
Then the vertex operators $\Lambda(\xi)$ and $\Omega(\xi)$
corresponding to single column and single row, respectively,
are defined by
\bea
\label{vertex-column}
&&\Lambda(\xi)
=
\vertex{\xi^{(1)}}{\xi^{(2)}}{\xi^{(s)}}{-1}
=
\prod_{\si=1}^s
      	\prod_{j=1}^{N^{(\si)}}
        (1-x^{(\si)}_j \xi^{(\si)}),
\\
\label{vertex-row}
&&\Omega(\xi)
=
\del{\mu}
\Gamma(\mu,\underbrace{\xi,\cdots,\xi}_{s-1};1/\alpha) |_{\mu=\xi}
=
\frac{1}{\alpha}
\sum_{n=1}^{\infty}p_n^{(1)}\xi^n
        \prod_{\si=1}^s
      	\prod_{j=1}^{N^{(\si)}}
        (1-x^{(\si)}_j \xi)^{-1/\alpha},
\eea
where $\del{\mu}=\mu\partial/\partial\mu$.
Notice that, in contrast to the spinless CSM,
the derivative in (\ref{vertex-row})
is essential for the case of single row.
It is easily to show that
\be
\label{single-column-H}
\calH(x)\Lambda =
\calHa(\xi)\Lambda,\qquad
\calHa
=
-(\sum_\si\del{\xi^{(\si)}})^2
+
(N+\alpha)\sum_\si\del{\xi^{(\si)}}
\ee
and
\be
\label{single-row-H}
\calH(x)\Omega =
\calHa(\xi)\Omega,\qquad
\calHa
=
\alpha\del{\xi}^2
+
(N-1)\del{\xi}.
\ee
The derivation of (\ref{single-column-H}) and (\ref{single-row-H})
is straightforward.
These formulae indicate that one particle and one hole excitations
of the spin CSM are reduced
to one-body problems in the ``dual'' system.
Later, we will prove the general version of the formula
(\ref{single-column-H}).

As mentioned above,
the states $\Lambda(\xi)$ and $\Omega(\xi)$ are generating functionals of
the eigenstates associated with the colored Young
diagram with single column $(1^n)$ and single row $(n)$, respectively.
Namely, by expanding these states in terms of $\xi^{(\si)}$ (or $\xi$),
\bea
\label{spin-jack-column}
&&
\Lambda(\xi)
=
\sum_{\si=1}^s\isum{n_\si}{0}{\infty}
\jack{n_1\cdots n_s}{-1}(x)
(\xi^{(1)})^{n_1}\cdots(\xi^{(s)})^{n_s},
\\
\label{spin-jack-row}
&&
\Omega(\xi)
=
\isum{n}{0}{\infty}\jack{n}{1/\alpha}(x)\xi^n,
\eea
we obtain eigenstates $\jack{n_1\cdots n_s}{-1}(x)$
and $\jack{n}{1/\alpha}(x)$.
Here $\jack{n_1\cdots n_s}{-1}(x)$ and $\jack{n}{1/\alpha}(x)$
denote the eigenstates corresponding to
the single column Young diagram
which has $n_\si$ boxes with color $\si$ and
the single row Young diagram with color $1$, respectively.

%
%
\section{Derivation of Dual Hamiltonian}
In this section, we calculate the action of the
Hamiltonian on product of the vertex operators
considered in the previous section.
We define the vertex operator $\Lambda(x|\xi)$,
\be
	\Lambda(x|\xi)
        =
        \prod_{\si=1}^s
      	\prod_{i=1}^{N^{(\si)}}
        \prod_{k=1}^{M^{(\si)}}
	(1-x^{(\si)}_i \xi^{(\si)}_k),
\label{dualHamil}
\ee
where $M^{(\si)}$ denotes the number of particles with spin
$\si$ in the dual system.
In what follows, we only consider the case such that
$|M^{(\si)}-M^{(\si')}|\leq1$ for all $\si, \si'$.
As in the previous section, the dual Hamiltonian $\hcH(\xi)$
is defined by
\be
	\cH(x)\Lambda(x|\xi)
	 = \hcH(\xi)\Lambda(x|\xi).
\ee
We decompose the Hamiltonian $\hcH$ as in \eq{decomp}.
Then the dual Hamiltonian $\hcH=
\sum_\si\hcH^{(\si)}(\xi)
+
\sum_{\si<\sj}\hcH^{(\si\sj)}_{int}(\xi)$
is given by,
\ba
\label{dualHamiltonian1}
\hcH^{(\si)}(\xi) & = &
-\sum_{k=1}^{M^{(\si)}}(D_{\xi_{k}^{(\si)}})^2
-
\alpha\sum_{k<l}
\frac{\xi_{k}^{(\si)}+\xi_{l}^{(\si)}}{\xi_{k}^{(\si)}-\xi_{l}^{(\si)}}
(D_{\xi_{k}^{(\si)}}-D_{\xi_{l}^{(\si)}}),
\\
\label{dualHamiltonian2}
\hcH^{(\si\sj)}_{int}(\xi) & = &
-2 \sum_{k=1}^{M^{(\si)}} \sum_{\ell = 1}^{M^{(\si')}}
\frac{
\prod_{s(\neq \ell)}^{M^{(\si')}} (1-\xi^{(\sj)}_s/\xi^{(\si)}_k)
\prod_{s(\neq k)}^{M^{(\si)}} (1-\xi^{(\si)}_s/\xi^{(\sj)}_\ell)
}{
\prod_{s(\neq k)}^{M^{(\si)}} (1-\xi^{(\si)}_s/\xi^{(\si)}_k)
\prod_{s(\neq \ell)}^{M^{(\si')}} (1-\xi^{(\sj)}_s/\xi^{(\sj)}_\ell)
}
D_{\xi^{(\si)}_k}D_{\xi^{(\sj)}_\ell}.
\ea
Here we omitted the terms which are proportional to
$\sum_k D_{\xi_{k}^{(\si)}}$.
Unlike the spinless CSM, the dual Hamiltonian is not similar to
the original one.
This fact reflects that the symmetry $\alpha
\leftrightarrow 1/\alpha$ is broken.

Notice that,
although this dual system does not described
by the ordinary two-body interaction,
its integrability is clear from our construction.
Moreover, we easily see that
it has the same spectrum as that of the original system.
In fact, if we expand $\Lambda $,
\be
\Lambda(x|\xi) =\sum_{\lambda} J_\lambda (x)
\hat{J_\lambda}(\xi),
\ee
where $J_\lambda(x)$ is the eigenstate of the original
Hamiltonian with
the colored diagram $\lambda=\{ \lambda_1\sigma_1,
\lambda_2\sigma_2,\cdots\}$,
then, because of \eq{dualHamil}, $\hat{J_\lambda}(\xi)$
should be the eigenstate of the dual Hamiltonian with the same
eigenvalue.


The derivation of the dual Hamiltonian is rather lengthy.
Then, for simplicity, we consider the case with
two components which we denote $\{\uparrow, \downarrow\}$.
Let $x_i$ and $y_i$ be the coordinates for the particles with
up and down spin, respectively,
and $\xi_k$ and $\eta_k$ be that of the dual system.
The derivation of eq.\ \eq{dualHamiltonian1} is straightforward.
To derive eq.\ \eq{dualHamiltonian2}, first we observe that
\be
\label{proof}
\cH_{int}^{(\uparrow\downarrow)}(x,y) \Lambda(x,y|\xi,\eta)
 =
\sum_{i,j}R(x_i,y_j)Q(x_i,y_j)\Lambda(x,y|\xi,\eta),
\ee
with
\ba
Q(x,y)
&=&
\prod_{k}\frac{1}{1-x \xi_k}
\prod_{\ell}\frac{1}{1-y \eta_\ell},
\non
R(x,y)
&=&
\frac{x+y}{x-y}\prod_k(1-x \xi_k)
\prod_\ell(1-y \eta_\ell)
\left(
\sum_k\frac{-x\xi_k}{1-x\xi_k}
-\sum_\ell\frac{-y\eta_\ell}{1-y\eta_\ell}
\right)\non
&&
-2\frac{xy}{(x-y)^2}
\left(
\prod_k(1-x\xi_k)\prod_\ell(1-y\eta_\ell)
 -\prod_k(1-y\xi_k)\prod_\ell(1-x\eta_\ell)
\right).\nonumber
\ea
Next we show that
the right hand side of the expression (\ref{proof}) can be
rewritten as the derivative with respect to $\xi$ and $\eta$
by combining following lemmas.

\begin{enumerate}
\item
We can rewrite $Q(x,y)$ as,
\be
Q(x,y) = \left( \sum_{k = 1}^\Mup A_k(\xi)
\frac{1}{1-x\xi_k}\right)
\left( \sum_{\ell = 1}^\Mdown A_\ell(\eta)
\frac{1}{1-y\eta_\ell}\right),
\ee
where $A_k(\xi)=\prod_{\ell(\neq k)}\frac{\xi_k}{\xi_k-\xi_\ell}$.
%
\item
$R(x,y)$ is a polynomial of degree $\Mup$ in $x$ and
that of degree $\Mdown$ in $y$.
Namely,
if we write,
\be
\prod_k (1-x\xi_k) = \sum_{n=0}^\Mup s_n(\xi) x^n,
\qquad
\prod_\ell (1-y\eta_\ell) = \sum_{m=0}^\Mdown s_m(\eta) y^m,
\ee
then $R(x,y)$ is expressed as
\be
R(x,y) = \sum_{n=0}^\Mup\sum_{m=0}^\Mdown s_n(\xi)
s_m(\eta) T_{n,m}(x,y),
\ee
where
\be
T_{n,m}(x,y)=
\left\{
\begin{array}{ll}
0,& n=m
\\
(n-m)x^ny^m
+2\sum_{r=1}^{n-m-1}(n-m-r)
x^{n-r}y^{m+r},& n>m
\\
(m-n)x^ny^m
+2\sum_{r=1}^{m-n-1}(m-n-r)
x^{n+r}y^{m-r},& n<m .
\end{array}
\right.
\ee
\item
For $0\leq n\leq \Mup$,
\be
\sum_{k=1}^\Mup A_k(\xi) \sum_{i=1}^\Nup
\frac{x_i^n}{1-x_i\xi_k}\Lambda
=
\left(
\delta_{n,0} \Ndown
-
\sum_{k=1}^{\Mup} A_k(\xi)
\xi_k^{-n} D_{\xi_k}
\right)
\Lambda
\ee
and the similar formula for $y$ and $\eta$ hold.
For the derivation of this formula,
we used the Euler's identity.
\end{enumerate}
By the first observation, the combination on the right hand
side of (\ref{proof}) can be expressed as derivative with respect
to $\xi$ and $\eta$
by using
$\partial_{\xi_k}\Lambda
=\sum_i\frac{-x_i}{1-x_i \xi_k}\Lambda$ etc.
The nontriviality comes from the $x,y$ dependence.
However, from the second observation, the dependence can
be reduced to their polynomial and then from
the third lemma they can
be replaced by the function of $\xi$ and $\eta$.
Therefore,
we finally obtain the interacting part of the dual Hamiltonian
$\hcH^{(\uparrow\downarrow)}_{int}(\xi,\eta)$.

\section{Discussions and Comments}
Although we know that the dual Hamiltonian
we derived is integrable, many of its properties are still
missing. One of such important issue is the existence
of the Hermitian measure.  If it exists, we can construct
every eigenstate of the spin CSM as we describe in the following.

Generalizing the spinless case \cite{AMOS,MY},
we define two transformations which map one eigenstate
into another.
The transformations are:

\begin{enumerate}
\item {\it Galilean transformation}: $G_P$\\
This transformation is defined by
\be
(G_P J)(x)
=
\left(\prod_{\si=1}^s\prod_{i=1}^{N^{(\si)}}
x^{(\si)}_i\right)^P J(x).
\ee
Since the spin CSM has the Galilean invariance, it obviously
maps one eigenstate to
another.
At the same time, the momentum of each particle is shifted
by $P$.  On the Young diagram, $G_P$ has an effect to
attach a rectangle Young diagram $(P^N)$
which has $N^{(\si)}$ rows with color $\si$'s.
This operation does not violate the rule of painting
and is always possible.

\item {\it Integral transformation which changes the
number of variables}: ${\cal N}(x,y)$\\
Let us denote the Hermitian inner product of the original
system as $\langle,\rangle_{x}$ and the inner product
for the dual system as $\langle\langle,\rangle\rangle_{\xi}$.
We define the integral transformation as,
\be
({\cal N}(x,y)J(x))(y)
=
\langle\langle
\Lambda(y|\xi),
\langle
\Lambda(x|\xi), J(x)
\rangle_{x}
\rangle\rangle_{\xi}.
\ee
If such inner product exists, from \eq{dualHamil},
it is clear that the Hamiltonian commutes with this operator
in a following sense:
$\cH(y){\cal N}(x,y) = {\cal N}(x,y) \cH(x)$.
Performing this transformation,
we can change the number of particles
for each color without touching the Young diagram.
\end{enumerate}

We can construct any eigenstate of the spin CSM
Hamiltonian
by alternate operations of these transformations to
the trivial eigenstate, namely the vacuum.

This construction of eigenstates is
a straightforward generalization of the method
which has been used in refs.\cite{AMOS,MY}
to obtain the integral representation
of the Jack polynomial,
and indicated the remarkable identification
between the Jack polynomial and
the singular vectors of the Virasoro
and $W_N$ algebras. 
We expect that the spin CSM also possesses such an algebraic structure.

Finally, we comment on the related topics.
The correspondence between
the eigenstates of the spin CSM and
the solutions of the Knizhnik-Zamolodchikov equation
has been established \cite{Hikamib,Que}.
More recently, Felder and Varchenko \cite{FB} (see also \cite{ES})
gave some formulae for the eigenstates of the spin CSM.
The Dunkl operators \cite{Dunkl} or
more precisely the representation theory
of the degenerate affine Hecke algebra \cite{Chere} have central r\^ole in
the analysis of the spectrum and integrability of the spin CSM
(see also \cite{LV}).
It would be interesting to clarify the relation between these works
and our results.
Also, it is natural to consider the $q$-analog of our methods.
The $q$-analog of the spin CSM has been constructed
\cite{Konno,Uglov}.
We hope to turn these issue in the near future.


\section*{Appendix: Examples of Eigenstates}

Here we give some explicit examples of the eigenstates of
the spin CSM Hamiltonian.
In the two components case,
the eigenstates $J_\lambda$ are written by two kinds of power sums
$\Pu n$ and $\Pd n$.
We distinguish between two colors of the Young diagram by
using bars, for example,
$J_{\lambda_1,\bar\lambda_2,\bar\lambda_3,\lambda_4,\cdots}$.
The eigenstates $J_\lambda$
with the Young diagrams $\lambda$ up to $3$ boxes are as follows:
$$
\Mat{J_1 \cr J_{\bar1}}
=
\Mat{\Pu1 \cr \Pd1},\qquad\quad
\Mat{J_2 \cr J_{\bar2} \cr J_{11} \cr J_{1\bar1} \cr J_{\bar2} }
=
\Mat{\a & 0  & 1 & 1 & 0 \cr
      0 &\a  & 0 & 1 & 1 \cr
     -1 & 0  & 1 & 0 & 0 \cr
      0 & 0  & 0 & 1 & 0 \cr
      0 &-1  & 0 & 0 & 1    }
\Mat{\Pu2 \cr \Pd2 \cr \Pu1\Pu1 \cr \Pu1 \Pd1 \cr \Pd1\Pd1 },
$$
$$
\Mat{J_3 \cr J_{\bar3} \cr
     J_{21} \cr J_{2\bar1} \cr J_{\bar21} \cr J_{\bar2\bar1} \cr
     J_{111} \cr J_{11\bar1} \cr J_{1\bar1\bar1} \cr J_{\bar1\bar1\bar1} }
=
\Mat{2\a^2 &   0    & 3\a  & 2\a  &  \a  &  0    & 1 & 2 & 1 & 0 \cr
       0   & 2\a^2  &  0   &  \a  & 2\a  & 3\a   & 0 & 1 & 2 & 1 \cr
     -\a   &   0    & \a-1 & -1   &  0   &  0    & 1 & 1 & 0 & 0 \cr
       0   &   0    &  0   & \a-1 &  1   &  0    & 0 & 1 & 1 & 0 \cr
       0   &   0    &  0   &  1   & \a-1 &  0    & 0 & 1 & 1 & 0 \cr
       0   & -\a    &  0   &  0   & -1   & \a-1  & 0 & 0 & 1 & 1 \cr
       2   &   0    & -3   &  0   &  0   &  0    & 0 & 0 & 0 & 0 \cr
       0   &   0    &  0   & -1   &  0   &  0    & 0 & 0 & 0 & 0 \cr
       0   &   0    &  0   &  0   & -1   &  0    & 0 & 0 & 0 & 0 \cr
       0   &   2    &  0   &  0   &  0   & -3    & 0 & 0 & 0 & 0    }
\Mat{\Pu3        \cr \Pd3        \cr
     \Pu2\Pu1    \cr \Pu2\Pd1    \cr \Pd2\Pu1    \cr \Pd2\Pd1    \cr
     \Pu1\Pu1\Pu1\cr \Pu1\Pu1\Pd1\cr \Pu1\Pd1\Pd1\cr \Pd1\Pd1\Pd1   }.
$$


Next we show some examples of the eigenstates 
of $N$ variables in the coordinate space.
The eigenstate and its eigenvalue
are parameterized by a non-negative sequence
$\lambda = (\lambda_1,\lambda_2,\cdots,\lambda_N \geq 0)$ and its set
$\{\lambda_1,\lambda_2,\cdots,\lambda_N\}$, respectively.
We define a monomial
$m_\lambda \equiv \prod_{i=1}^N x_i^{\lambda_i}$.
The first few examples of the eigenstates $J_\lambda$,
in the case of $\sum_{i=1}^N \lambda_i = N$,
are as follows:
$$
J_1 = m_1,\qquad
J_{20} = (\a+1) m_{20} + m_{11}, \qquad
J_{11} = m_{11},
$$
$$
J_{300} =
(\a+1)(2\a+1) m_{300}
+ (\a+1) \left(2m_{210} + 2m_{201} + m_{120} + m_{102}\right) + 2m_{111},
$$
$$
J_{210} =
(\a+2) m_{210} + m_{111},\qquad
J_{111} = m_{111}.
$$
Eigenstates $J_{02}$, $J_{030}$ and $J_{201}$ etc. have the similar forms.
Remark that if we sum up $J_\lambda$
over $\lambda$'s which have the same set $\{\lambda\}$,
then we obtain the Jack polynomial
with the corresponding Young diagram $\{\lambda\}$.

\vskip 2mm
{\bf Acknowledgments}
\vskip 2mm
\noindent

Our thanks are due to S.~Nam and O.~Tsuchiya
for discussions in the earlier stage of this work.
We would like to thank
Y.~Kato, A.N.~Kirillov, S.~Odake and D.~Uglov
for discussions.
%
H.A. was partly supported by
a Grant-in-Aid for Scientific Research from the Ministry of Education,
Science and Culture, Japan.
The work of Y.M. was supported in part by
Grant-in-Aid for Scientific Research
on Priority Area 231 ``Infinite Analysis'',
the Ministry of Education, Science and Culture, Japan.
T.Y. was supported by
the COE (Center of Excellence) researchers program
of the Ministry of Education, Science and Culture, Japan.



\begin{thebibliography}{99}

\bibitem{C}
F.~Calogero,
J. Math. Phys. {\bf 10} (1969) 2197.

\bibitem{S}
B.~Sutherland,
J. Math. Phys. {\bf 12} (1970) 246, 251,
Phys. Rev. A{\bf 4} (1971) 2019, A{\bf 5} (1992) 1372.

\bibitem{EX}
F.~D.~M.~Haldane,
Phys. Rev. Lett. {\bf 66} (1991) 1529.

\bibitem{BW}
D.~Bernard and Y.-S.~Wu,
in
{\it New Developments of Integrable Systems and Long-Ranged
Interaction Models}, eds. M.-L. Ge and Y.-S. Wu
(World Scientific 1995).
%
\bibitem{Ha}
Z.~N.~C.~Ha,
Nucl. Phys. {\bf B435} (1995) 604;
F.~Lesage, V.~Pasquier and D.~Servan,
Nucl. Phys. {\bf B435} (1995) 585.

\bibitem{revCS}
For reviews, see following papers and references therein:\\
N.~Kawakami,
Prog. Theor. Phys. {\bf 91} (1994) 189;\\
F.~D.~M.~Haldane,
in
{\em Correlation Effects in Low Dimensional Electron Systems},
eds. A.~Okiji and N.~Kawakami (Springer, 1994).




\bibitem{MIT}
B.~D.~Simons, P.~A.~Lee and B.~L.~Altshuler,
Nucl. Phys. {\bf B409} (1993) 487.

\bibitem{HS}
 F.~D.~M.~Haldane,
Phys. Rev. Lett. {\bf 60} (1988) 635;\\
B.~S.~Shastry,
Phys. Rev. Lett. {\bf 60} (1988) 639.

\bibitem{HaHa}
Z.~N.~C.~Ha and F.~D.~M.~Haldane,
Phys. Rev. B{\bf 46} (1992) 9359.

\bibitem{KaKu}
Y.~Kato and Y.~Kuramoto,
Phys. Rev. Lett. {\bf 74} (1995) 1222.
%
\bibitem{MP}
J.~A.~Minahan and A.~P.~Polychronakos,
Phys. Lett. B {\bf 302} (1993) 265.
%
\bibitem{HW}
K.~Hikami and M.~Wadati,
J. Phys. Soc. Jpn. {\bf 62} (1993) 4203.

\bibitem{HHTBP}
F.~D.~M.~Haldane,  Z.~N.~C.~Ha, J.~.C~.~Talstra,
D.~Bernard and V.~Pasquier,
Phys. Rev. Lett. {\bf 69} (1992) 2021.

\bibitem{BGHP}
D.~Bernard, M.~Gaudin, F.~D.~M.~Haldane and V.~Pasquier,
J. Phys. A {\bf 26} (1993) 5219.

\bibitem{Hikamia}
K.~Hikami,
J. Phys. A {\bf 28} (1995) L131.
%
\bibitem{BHW}
D.~Bernard, K.~Hikami and M.~Wadati,
in
{\em New Developments of Integrable Systems and
Long-Ranged Interaction Models},
eds. M.-L. Ge and Y.-S. Wu
(World Scientific, 1995).

\bibitem{BPS}
D.~Bernard, V.~Pasquier and D.~Serban,
Nucl. Phys. {\bf B428} (1994) 612.
%
\bibitem{BLS}
P.~Bouwknegt, A.~W.~W.~Ludwig and K.~Schoutens,
Phys. Lett. B {\bf 338} (1994) 448.
%
\bibitem{AN}
C.~Ahn and S.~Nam,
{\it Yangian Symmetries in the $SU(N)_1$ WZW Model and the
Calogero-Sutherland Model},
preprint (1995) SNUTP/9, hep-th/9510241.


\bibitem{AMOS}
H.~Awata, Y.~Matsuo, S.~Odake and J.~Shiraishi,
Phys. Lett. B {\bf 347} (1995) 49,
Nucl. Phys. {\bf B449} (1995) 347.

\bibitem{MY}
K.~Mimachi and Y.~Yamada,
  {\it Singular vectors of the Virasoro algebra in terms of
  Jack symmetric polynomials},
preprint (1994),
to appear in Commun. Math. Phys.,\\
RIMS {\it Kokyuroku} {\bf 919} (1995) 68 ({\it in Japanese}).
%
\bibitem{Iso}
S.~Iso,
Nucl. Phys. {\bf B443} (1995) 581.
%
\bibitem{MPb}
J.~Minahan and A.~P.~Polychronakos,
Phys. Rev. B{\bf 50} (1994) 4236.
%
\bibitem{Khv}
D.~V.~Khveshchenko,
Int. J. Mod. Phys. B {\bf 9} (1995) 1639.
%


%
\bibitem{Hikamib}
K.~Hikami,
J. Phys. A {\bf 27} (1994) L541.
%
\bibitem{Que}
C.~Quesne,
J. Phys. A {\bf 28} (1995) 3533.
%
\bibitem{FB}
G.~Felder and A.~Varchenko,
{\it Three formulas for eigenfunctions
of integrable Schroedinger operators},
preprint (1995) hep-th/9511120.


\bibitem{ES}
P.~Etingof and K.~Styrkas,
Compos. Math. {\bf 98} (1995) 91.


\bibitem{Dunkl}
C.~F.~Dunkl,
Trans. AMS {\bf 311} (1989) 167.

\bibitem{LV}
L.~Lapointe and L.~Vinet,
{\it Exact Operator Solution of the Calogero-Sutherland model Author},
preprint (1995) hep-th/9507073.


\bibitem{Chere}
I.~Cherednik,
Adv. Math. {\bf 106} (1994) 65.
%


\bibitem{Konno}
H.~Konno,
{\it Relativistic Calogero-Sutherland Model: Spin Generalization, Quantum
Affine Symmetry and Dynamical Correlation Functions},
preprint (1995) YITP/K-1118, hep-th/9508016.
%
\bibitem{Uglov}
D.~Uglov,
{\it The trigonometric counterpart of the Haldane Shastry model},
preprint (1995) hep-th/9508145.

%
%





\end{thebibliography}
\end{document}